\documentclass[12pt]{article}
\usepackage{amssymb,amsmath}
\usepackage{graphicx}

\newcommand{\be}{\begin{equation}}
\newcommand{\ee}{\end{equation}}
\newcommand{\bea}{\begin{eqnarray}}
\newcommand{\eea}{\end{eqnarray}}
\newcommand{\beas}{\begin{eqnarray*}}
\newcommand{\eeas}{\end{eqnarray*}}
\newcommand{\ba}{\begin{array}}
\newcommand{\ea}{\end{array}}

\newcommand{\tr}{\mathrm{Tr}}

\newcommand{\nbox}{{\,\lower0.9pt\vbox{\hrule \hbox{\vrule height 0.2 cm \hskip 0.19 cm \vrule height 0.2 cm}\hrule}\,}}

\def\href#1#2{#2}

\begin{document}
\begin{titlepage}
\hfill
\vbox{
    \halign{#\hfil         \cr
           } 
      }  
\vspace*{20mm}
\begin{center}
{\Large \bf The Gravity Dual of a Density Matrix}

\vspace*{15mm}
\vspace*{1mm}
Bart{\l}omiej Czech, Joanna L.~Karczmarek, Fernando Nogueira, Mark Van Raamsdonk
\vspace*{1cm}

{
Department of Physics and Astronomy,
University of British Columbia\\
6224 Agricultural Road,
Vancouver, B.C., V6T 1Z1, Canada}

\vspace*{1cm}
\end{center}

\begin{abstract}
For a state in a quantum field theory on some spacetime, we can associate a density matrix to any subset of a given spacelike slice by tracing out the remaining degrees of freedom. In the context of the AdS/CFT correspondence, if the original state has a dual bulk spacetime with a good classical description, it is natural to ask how much information about the bulk spacetime is carried by the density matrix for such a subset of field theory degrees of freedom. In this note, we provide several constraints on the largest region that can be fully reconstructed, and discuss specific proposals for the geometric construction of this dual region.

\end{abstract}

\end{titlepage}

\vskip 1cm

\section{Introduction}

The AdS/CFT correspondence \cite{malda, agmoo} relates states of a field theory on some fixed spacetime {\bf B} to states of a quantum gravity theory for which the spacetime metric is asymptotically locally AdS with boundary geometry {\bf B}. The field theory provides a nonperturbative description of the quantum gravity theory that is manifestly local on the boundary spacetime: for a given spacelike slice of the boundary spacetime {\bf B}, the degrees of freedom in one subset are independent from the degrees of freedom in another subset. On the gravity side, identifying independent degrees of freedom is much more difficult; for example, the idea of black hole complementarity \cite{stu} suggests that local excitations inside the horizon of a black hole cannot be independent of the physics outside the horizon. It is therefore interesting to ask whether we can use our knowledge of independent field theory degrees of freedom to learn anything about which degrees of freedom on the gravity side may be considered to be independent.

In this paper, we consider the following question: Given a CFT on {\bf B} in a state $|\Psi \rangle$ dual to a spacetime $M$ with a geometrical description, and given a subset $A$ of a spatial slice of {\bf B}, what part of the spacetime $M$ can be fully reconstructed from the density matrix $\rho_A$ describing the state of the subset of the field theory degrees of freedom in $A$?

An immediate question is why we expect there to be any region that can be reconstructed if we know only about the degrees of freedom on a subset of the boundary. If the map between boundary degrees of freedom and the bulk spacetime is sufficiently non-local, it could be that information from every region of the boundary spacetime is needed to reconstruct any particular subset of $M$. However, there are various reasons to be more optimistic. It is well known that the asymptotic behavior of the fields in the bulk spacetime is given directly in terms of expectation values of local operators in the field theory (together with the field theory action). Equipped with this boundary behavior of the bulk fields in some region of the boundary\footnote{As we recall below, knowledge of the field theory density matrix for a spatial region $A$ allows us to compute any field theory quantities localized to a particular codimension-zero region of the boundary, the domain of dependence of $A$.} and the bulk field equations, we should be able to integrate these field equations to find the fields in some bulk neighborhood of this boundary region. We can also compute various other field theory quantities (e.g. correlation functions, Wilson loops, entanglement entropies) restricted to the region $A$ or its domain of dependence. According to the AdS/CFT dictionary, these give us direct information about nearby regions of the bulk geometry.

The notion that particular density matrices can be associated with certain patches of spacetime was advocated in \cite{VanRaamsdonk:2009ar}.\footnote{For an earlier discussion of mixed states in the context of AdS/CFT, see \cite{fhmmrs}.} There, it was pointed out that a given density matrix may arise from many different states of the full system, or from a variety of different quantum systems that contain this set of degrees of freedom as a subset. Different pure states that give rise to the same density matrix for the subset correspond to different spacetimes with a region in common; this common region can be considered to be the dual of the density matrix.\footnote{As a particular example, it was pointed out in \cite{VanRaamsdonk:2009ar} that a CFT on $S^d$ in a thermal density matrix, commonly understood to be dual to an AdS/Schwarzchild black hole, cannot possibly know whether the whole spacetime is the maximally extended black hole; only the region outside the horizon is common to all states of larger systems for which the CFT on $S^d$ forms a subset of degrees of freedom described by a thermal density matrix.}

In the bulk of this paper, we seek to understand in general the region of a bulk spacetime $M$ that can be directly associated with the density matrix describing a particular subset of the field theory degrees of freedom. We begin in Section~2 by reviewing some relevant facts from field theory and arguing that the density matrix associated with a region $A$ may be more naturally associated with the domain of dependence $D_A$ (defined below). In Section~3, we outline in more detail the basic question considered in the paper. In Section~4, we propose several basic constraints on the region $R(A)$ dual to a density matrix $\rho_A$. In Section~5, we consider two regions that are plausibly contained in $R(A)$. First, we argue that $z(D_A)$, the intersection of the causal past and causal future of $D_A$, satisfies our constraints and should be contained in $R(A)$, as should its domain of dependence, $\hat{z}(D_A)$.\footnote{We denote domains of dependence in the boundary with $D_\cdot$ (for example, $D_A$), while domains of dependence in the bulk are marked with a hat $\hat{\phantom{.}}$.} We note that in some special cases, $R(A)$ cannot be larger than $\hat z(D_A)$. However, in generic spacetimes, we argue that entanglement observables that can be calculated from the density matrix $\rho_A$ certainly allow us to probe regions of spacetime beyond $\hat
z(D_A)$.\footnote{It is an open question whether these observables are enough to reconstruct the spacetime beyond $\hat z(D_A)$, so we cannot say with certainty that $R(A)$ is larger than $\hat{z}(D_A)$.} This motivates us to consider another region, $w(D_A)$, defined as the union of surfaces used to calculate these entanglement observables (defined more precisely below) according to the holographic entanglement entropy proposal \cite{Ryu:2006bv, Hubeny:2007xt}. We show that $w(D_A)$ (or more precisely, its domain of dependence $\hat{w}(D_A)$) also satisfies our constraints, and that for a rather general class of spacetimes, there is a sense in which $R(A)$ cannot be larger than $\hat{w}(D_A)$. On the other hand, we show that in some examples, $R(A)$ must be larger than $\hat{w}(D_A)$. We conclude in Section~6 with a summary and discussion.
\\
\\
{\bf Note added:} While this paper was in preparation, \cite{Bousso:2012sj} appeared, which has some overlap with our discussion. We also became aware of \cite{HRnew}, which considers related questions.

\section{Field theory considerations}

To begin, consider a field theory on some globally hyperbolic spacetime {\bf B}, and consider a spacelike slice $\Sigma$ that forms a Cauchy surface. Then, classically, the fields on this hypersurface and their derivatives with respect to some timelike future-directed unit vector orthogonal to the hypersurface determine the complete future evolution of the field. Quantum mechanically, the fields on this hypersurface can be taken as the basic set of variables for quantization and conjugate momenta defined with respect to the timelike normal vector.

Now consider some region $A$ of the hypersurface $\Sigma$. Since the field theory is local, the degrees of freedom in $A$ are independent from the degrees of freedom in the complement $\bar{A}$ of $A$ on $\Sigma$. Thus, the Hilbert space can be decomposed as a tensor product ${\cal H} = {\cal H}_A \otimes {\cal H}_{\bar{A}}$, and we can associate a density matrix
$\rho_A = \tr_{\bar{A}}(|\Psi \rangle \langle \Psi \rangle)$ to the degrees of freedom in $A$. This density matrix captures all information about the state of the degrees of freedom in $A$ and can be used to compute any observables localized to $A$.

\begin{figure}
\centering
\includegraphics[width=0.3\textwidth]{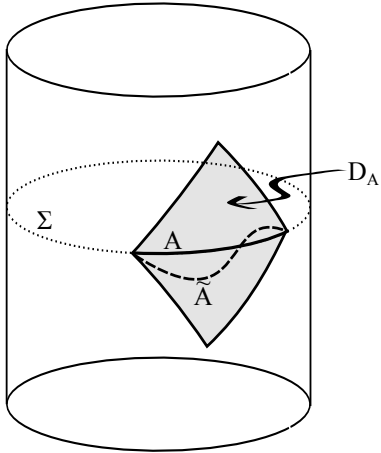}
\caption{A spacelike slice $\Sigma$ of a boundary manifold {\bf B} ($ = S^1 \times$ time) with a region $A$ and its domain of dependence $D_A$. The same domain of dependence arises from any spacelike boundary region $\tilde{A}$ homologous to $A$ with $\partial A = \partial \tilde{A}$.}
\label{domaind}
\end{figure}

In fact, the density matrix $\rho_A$ allows us to compute field theory observables localized to a larger region $D_A$ known as the domain of dependence of $A$. The domain of dependence $D_A$ is the set of points $p$ in {\bf B} for which every (inextensible) causal curve through $p$ intersects $A$ (see Figure~\ref{domaind}). Classically, the region $D_A$ is the subspace of {\bf B} in which the field values are completely determined in terms of the initial data on $A$. Quantum mechanically, any operator in $D_A$ can be expressed in terms of the fields in $A$ alone and therefore computed using the density matrix $\rho_A$.

As can be seen from Figure~\ref{domaind}, any other spacelike surface $\tilde{A}$ homologous to $A$ with boundary $\partial \tilde{A} = \partial A$ shares its domain of dependence.\footnote{To see this, we note that since $A$ and $\tilde{A}$ are homologous, we can deform $A$ into $\tilde{A}$ and define $\cal B$ to be the volume bound by $A$ and $\tilde{A}$. Then for any point $p$ in ${\cal B}$, consider an inextensible causal curve through $p$. Such a curve must necessarily pass through $A$. But it cannot pass through $A$ twice, since $A$ is spacelike. On the other hand, the curve must intersect the boundary of the region $\cal B$ twice (on the past boundary and on the future boundary), so it must have an intersection with $\tilde{A}$.}  Thus, in some other quantization of the theory based on a hypersurface $\tilde{\Sigma}$ with $\tilde{A} \subset \tilde{\Sigma}$, we expect that the density matrix $\rho_{\tilde{A}}$ contains the same information as the density matrix ${\rho_A}$. It is then perhaps more natural to associate density matrices directly with domain of dependence regions. This observation is important for our considerations below: in constructing the bulk region dual to a density matrix $\rho_A$, it is more natural to use the boundary region $D_A$ as a starting point, rather than the surface $A$.

It is useful to note that a quantum field theory on a particular domain of dependence can be thought of as a complete quantum system, independent of the remaining degrees of freedom of the field theory. The observables of this field theory are the set of all operators built from the fields on $A$. The state of the theory is specified by a density matrix $\rho_A$, which allows us to compute any such observable. The spectrum of this density matrix, and associated observables such as the von Neumann entropy, give additional information about the system. We can interpret this in a thermodynamic way as giving information about the ensemble of pure states described by the density matrix. Alternatively, viewing this system as a subset of a larger system that we assume is in a pure state, we can interpret this additional information as telling us about the entanglement between the degrees of freedom in our causal development region with other parts of the system.

\section{The gravity dual of $\rho_A$}

In this section, we consider the question of how much information the density matrix $\rho_A$ carries about the dual spacetime. We restrict the discussion to states of the full system that are dual to some spacetime M with a good classical description.
Specifically, we ask the question
\vskip 0.1 in
\noindent
{\bf Question:}  {\it Suppose that a field theory on a spacetime {\bf B} in a state $|\Psi \rangle$ has a dual spacetime $M$ with a good geometrical description (e.g. a solution to some low-energy supergravity equations). How much of $M$ can be reconstructed given only the density matrix $\rho_A$ for the degrees of freedom in a subset $A$ of some spacelike slice of the boundary?}
\vskip 0.1 in
\noindent
Alternatively, we can ask:
\vskip 0.1 in
\noindent
{\it Consider all states $|\Psi_\alpha \rangle$ with dual spacetimes $M_\alpha$ that give rise to a particular density matrix $\rho_A$ for region $A$ of the boundary spacetime. What is the largest region common to all the $M_\alpha$s? }
\vskip 0.1 in
\noindent
We recall that knowledge of the density matrix $\rho_A$ allows us to calculate any field theory observable involving operators localized in the domain of dependence $D_A$, plus additional quantities such as the entanglement entropy associated with the degrees of freedom on any subset of $A$. According to the AdS/CFT dictionary, these observables give us a large amount of information about the bulk spacetime, particularly near the boundary region $D_A$, so it is plausible that at least some region of the bulk spacetime can be fully reconstructed from this data. We will refer to this region as $R(A)$.  We expect that in general the density matrix $\rho_A$ carries additional information about some larger region $G(A)$, but this additional information does not represent the complete information about $G(A) - R(A)$.

In this paper, we do not attempt to come up with a procedure to reconstruct the region $R(A)$; rather we will attempt to use general arguments to constrain how large $R(A)$ can be.

\section{Constraints on the region dual to $\rho_A$}

Before considering specific proposals for $R(A)$, it will be useful to point out various constraints that $R(A)$ should satisfy. First, since the density matrices for any two subsets $A$ and $\tilde A$ with the same domain of dependence $D$ correspond to the same information in the field theory, we expect that the region of spacetime that can be reconstructed from $\rho_A$ is the same as the region that can be reconstructed from $\rho_{\tilde A}$. Thus we have:
\vskip 0.1 in
\noindent
{\bf Constraint 1:}  {\it If $A$ and $\tilde A$ have the same domain of dependence $D$, then $R(A)= R(\tilde A)$.}
\vskip 0.1 in
\noindent
For a particular boundary field theory, the bulk spacetime will be governed by some specific low-energy field equations. We assume that we are working with a known example of AdS/CFT so that these equations are known. If we know all the fields in some region $R$ of the bulk spacetime $M$, we can use these field equations to find the fields everywhere in the bulk domain of dependence of $R$ (which we denote by $\hat{R}$). Since $R(A)$ is defined to be the largest region of the bulk spacetime that we can reconstruct from $\rho_A$, we must have:
\vskip 0.1 in
\noindent
{\bf Constraint 2:}  {\it $\hat{R}(A) = R(A)$.}
\vskip 0.1 in
\noindent
Now, suppose we consider two non-intersecting regions $A$ and $B$ on some spacelike slice of the boundary spacetime. The degrees of freedom in $A$ and $B$ are completely independent, so it is possible to change the state $|\Psi \rangle$ such that $\rho_B$ changes but $\rho_A$ does not.\footnote{Further, we expect that for some subset of these variations, the dual spacetime continues to have a classical geometric description.} Changes in $\rho_B$ will generally affect the region $R(B)$ in the bulk spacetime, but as a consequence can also affect any region in the causal future $J^+(R(B))$ or causal past $J^-(R(B))$ of $R(B)$. But these changes can have no effect on the region $R(A)$ since this region can be reconstructed from $\rho_A$, which does not change. Thus, we have:
\vskip 0.1 in
\noindent
{\bf Constraint 3:}  {\it If $A$ and $B$ are non-intersecting regions of a spacelike slice of the boundary spacetime, then $R(A)$ cannot intersect $J(R(B))$.}
\vskip 0.1 in
\noindent
Here we have defined $J(R) = J^-(R) \cup J^+(R)$. Note that whatever $R(B)$ is, it certainly includes $D_B$ so as a corollary, we can say that $R(A)$ cannot intersect $J(D_B)$. Taking $B = \bar{A}$ (i.e. as large as possible without intersecting $A$), we get a definite upper bound on the size of $R(A)$: it cannot be larger than the complement of $J(D_{\bar{A}})$.

\section{Possibilities for $R(A)$}

Let us now consider some physically motivated possibilities for the region $R(A)$. An optimistic expectation is that we could reconstruct the entire region $G(A)$ of the bulk spacetime $M$ used in calculating any field theory observable localized in $D_A$ (for example, all points touched by any geodesic with boundary points in $D_A$). However, this cannot be a candidate for $R(A)$, since it is easy to find examples of non-intersecting $A$ and $B$ on some spacelike slice of a boundary spacetime such that geodesics with endpoints in $B$ intersect with geodesics with endpoints in $A$.\footnote{For example, suppose we consider the vacuum state of a CFT on a cylinder and take $A$ and $B$ to be the regions $\theta \in (0,\pi/2)\cup(\pi,3 \pi/2)$ and $\theta \in (\pi/2,\pi) \cup (3\pi/2 , 2\pi)$ on the $\tau=0$ slice. Then the lines of constant $\theta$ are spatial geodesics in the bulk, and the region covered by such geodesics anchored in $A$ clearly intersects the region of such geodesics anchored in $B$.} Thus, $G(A) \cap G(B) \ne \emptyset$ (which implies $G(A) \cap J(G(B)) \ne \emptyset$) and so Constraint 3 is violated.

A lesson here is that even if field theory observables calculated from a boundary region $D_A$ probe a certain region of the bulk, they cannot necessarily be used to reconstruct that region. Generally, we will have $R(A) \subset G(A) \subset M$, where $\rho_A$ contains complete information about $R(A)$, some information about $G(A)$ and no information about $\bar{G}(A)$.

\subsection{The causal wedge $z(D_A)$}

\begin{figure}
\centering
\includegraphics[width=0.24\textwidth]{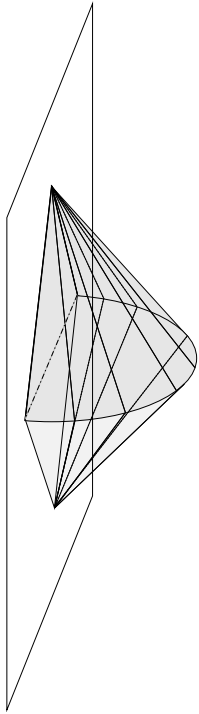}
\caption{Causal wedge $z(D_A)$ associated with a domain of dependence $D_A$.}
\label{zd}
\end{figure}

A simple region that is quite plausibly included in $R(A)$ is the set of points $z(D_A)$ in the bulk that a boundary observer restricted to $D_A$ can communicate with (i.e. send a light signal to and receive a signal back). For example, such an observer could easily detect the presence or absence of an arbitrarily small mirror placed at any point in $z(D_A)$. Formally, this region in the bulk is defined as the intersection of the causal past of $D_A$ with the causal future of $D_A$ in the bulk, $z(D_A) \equiv J^+(D_A) \cap J^-(D_A)$, as shown in Figure~\ref{zd}.\footnote{Recall that the causal future $J^+(D_A)$  of $D_A$ in the bulk is the set of points reachable by causal curves starting in $D_A$ while the causal past $J^-(D_A)$ of $D_A$ is the set of points, from which $D_A$ can be reached along a causal curve.} These observations correspond to perturbing the spacetime at one point in the asymptotic region and observing the asymptotic fields at another point at a later time. In the field theory language, such observations correspond to calculating response functions, in which the fields are perturbed at one point in $D_A$ and observed at another point in $D_A$. Such calculations can be done using only the density matrix $\rho_A$, thus we expect that $z(D_A)$ is included in the region $R(A)$.

By condition 2, we can extend this expectation to the proposal that $\hat{z}(D_A) \subset R(A)$. It is straightforward to check that $\hat{z}(D_A)$ also satisfies condition 3.\footnote{Suppose subsets $A$ and $B$ of a boundary slice do not intersect and suppose $p \in J(\hat{z}(D_B))$. Then there exists a causal curve through $p$ that intersects $\hat{z}(D_B)$ and therefore intersects some $q$ in $z(D_B)$. If $p$ is also in $\hat{z}(D_A)$, this same causal curve through $p$ must intersect a point $r$ in $z(D_A)$. Thus, there is a causal curve from $q$ in $z(D_B)$ to $r$ in $z(D_A)$. By definition of $z$, we must be able to extend this curve to a causal curve connecting $D_A$ to $D_B$. But in this situation, perturbations to the fields in $D_A$ could affect the fields in $D_B$ (or vice versa), and this would violate field theory causality.} Thus, the suggestion that $\hat{z}(D_A) \subset R(A)$ is consistent with our Constraints 1, 2  and 3.

The boundary of the region $z(D_A)$ in the interior of the spacetime is a horizon with respect to the boundary region $D_A$. Thus, the statement that we can reconstruct the region $z(D_A)$ is equivalent to saying that the information in $D_A$ is enough to reconstruct the spacetime outside this horizon. This horizon can be an event horizon for a black hole, but in general is simply a horizon for observers restricted to the boundary region $D_A$.

\begin{figure}
\centering
$\begin{array}{lcr}
\includegraphics[width=0.25\textwidth]{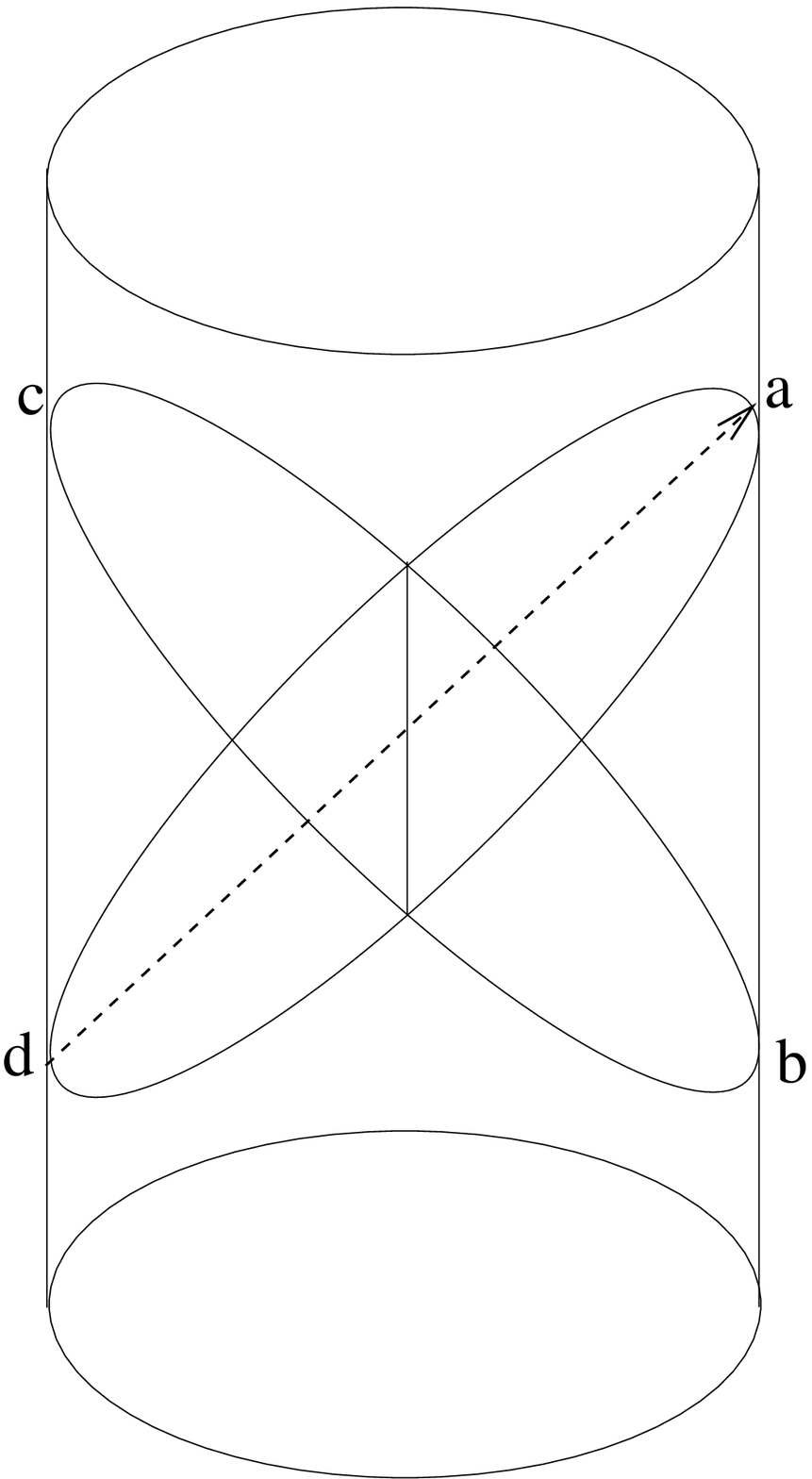}
& \qquad \qquad \qquad \qquad &\includegraphics[width=0.25\textwidth]{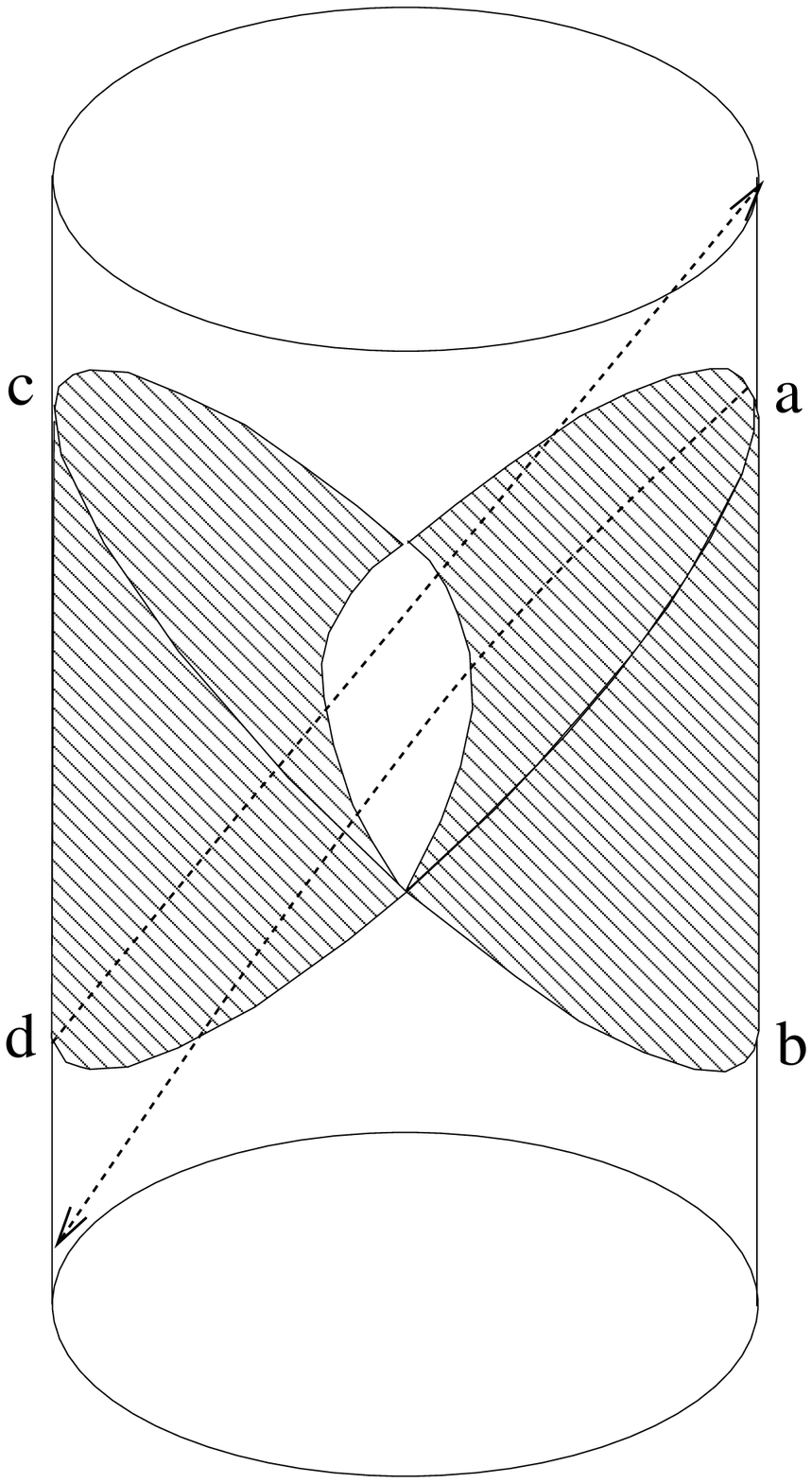}
\end{array}$
\caption{In pure global AdS, causal wedges of complementary hemispherical regions of the $\tau=0$ slice intersect along a codimension-two surface. In generic asymptotically AdS spacetimes, they intersect only at the boundary.}
\label{wedgeshole}
\end{figure}

In certain simple examples, it is straightforward to argue that $R(A)$ cannot be larger than $z(D_A)$ or $\hat{z}(D_A)$. For example, if $M$ is pure global AdS spacetime and $A$ is a hemisphere of the $\tau=0$ slice of the boundary cylinder, then $z(D_A)$ is the region bounded by the lightcones from the past and future tips of $D_A$ and the spacetime boundary, as shown in Figure \ref{wedgeshole}. Any point outside this region is in the causal future or causal past of the boundary region $D_{\bar{A}}$,\footnote{This relies on the fact that for pure global AdS, the forward lightcone from the past tip of $D_A$ (point b in Figure \ref{wedgeshole}) is the same as the backward lightcone from the future tip of $D_{\bar{A}}$ (point c) and the backward lightcone from the future tip of $D_A$ (point a in Figure \ref{wedgeshole}) is the same as the forward lightcone from the past tip of $D_{\bar{A}}$ (point d).} so by Constraint 3 (and the consequences discussed afterwards) such points cannot be in $R(A)$.

\subsubsection*{Information beyond the causal wedge $z(D_A)$}

We might be tempted to guess that $R(A)=\hat{z}(D_A)$ in general, but we will now see that $\rho_A$ typically contains significant information about the spacetime outside the region $\hat{z}(D_A)$.
Consider the same example of a CFT on the cylinder with the same regions $A$ and $\bar{A}$, but now consider some other state for which the dual spacetime is not pure AdS. A key observation\footnote{We are grateful to Veronika Hubeny and Mukund Rangamani for pointing this out.} is that, generically, the wedges $z(D_A)$ and $z(D_{\bar{A}})$ do not intersect, except at the boundary of $A$. This follows from a result of Gao and Wald \cite{Gao:2000ga} that light rays through the bulk of a generic asymptotically AdS spacetime generally take longer to reach the antipodal point of the sphere than light rays along the boundary. Thus, the backward lightcone from the point $a$ in the right panel of Figure~\ref{wedgeshole} is different from the forward lightcone from point $d$. We can still argue that $R(A)$ cannot overlap with the region
$J^+(D_{\bar A}) \cup J^-(D_{\bar A })$, but the complement of this region no longer coincides with $\hat z(D_A)$. Thus, it is possible that $R(A)$ is larger than $\hat z(D_A)$ in these general cases.\footnote{As an explicit example of a spacetime with this property, we can take a static spacetime with a spherically symmetric configuration of ordinary matter in the interior, e.g. the boson stars studied in \cite{Astefanesei:2003qy}.}

To see that the density matrix $\rho_A$ typically does contain information about the spacetime outside the region $z(D_A)$, we can take inspiration from Hubeny \cite{Hubeny:2012ry}, who argued that in many examples, the field theory observables that probe deepest into the bulk of the spacetime are those computed by extremal codimension-one surfaces in the bulk.

According to the proposal of Ryu and Takayanagi \cite{Ryu:2006bv} and the covariant generalization by Hubeny, Rangamani, and Takayanagi \cite{Hubeny:2007xt}, the von Neumann entropy of a density matrix $\rho_C$ corresponding to any spatial region $C$ of the boundary gives the area of a surface $W(C)$ in the bulk defined either as
\begin{itemize}
\item
the extremal codimension-two surface $W$ in the bulk whose boundary is the boundary of $C$. In cases where more than one such extremal surface exists, we take the one with minimal area, or
\item
the surface of minimal area such that the light sheets from this surface intersect the boundary exactly at $\partial D_C.$
\end{itemize}
In each case, it is assumed that the surface $W$ is homologous to $C$. In \cite{Hubeny:2007xt}, it is argued that these two definitions are equivalent.

Now, consider the surface $W(A)$ that computes the entanglement entropy of the full density matrix $\rho_A$. From the second definition, it is clear that the surface $W$ cannot have any part in the interior of $z(D_A)$. Otherwise, the light rays from any such point would reach the boundary in the interior of region $D_A$, and it would not be true that the light sheet from $W$ intersects the boundary at $\partial D_A$. By the same argument, the surface $W(\bar{A})$ that computes the entanglement entropy of $\rho_{\bar{A}}$ cannot have any part in the interior of $z(D_{\bar{A}})$. But by the first definition, the surface $W(\bar{A})$ is the same as the surface $W(A)$, since $\partial \bar{A} = \partial A$.\footnote{The equivalence of these surfaces and hence their areas is consistent with the fact that for a pure state in a Hilbert space ${\cal H} = {\cal H}_A \otimes {\cal H}_{\bar{A}}$, the spectrum of eigenvalues of $\rho_A$ must equal the spectrum of eigenvalues of $\rho_{\bar{A}}$. Thus, the entanglement entropies $S(\rho_A)$ and $S(\rho_{\bar{A}})$ must agree. We do not consider here the case where the entire theory is in a mixed state.} Since $z(D_A)$ and $z(D_{\bar{A}})$ generally have no overlap in the bulk of the spacetime, it is now clear that the surface $W$ lies outside at least one of $z(D_A)$ and $z(D_{\bar{A}})$.

To summarize, the area of surface $W$ may be computed as the von Neumann entropy of either the density matrix $\rho_A$ or the density matrix $\rho_{\bar{A}}$. In the generic case where $z(D_A)$ and $z(D_{\bar{A}})$ do not intersect in the bulk, the surface $W$ must lie outside at least one of $z(D_A)$ and $z(D_{\bar{A}})$. Thus, we can say that either the density matrix $\rho_A$ carries some information about the spacetime outside $z(D_A)$ or the density matrix $\rho_{\bar{A}}$ carries information about the spacetime outside $z(D_{\bar A})$.\footnote{Again, it is easy to check this in specific examples. For explicit examples of spherically symmetric static star geometries asymptotic to global AdS with $A$ equal to a hemisphere of the $\tau=0$ slice, the surface $W(A)$ lies at $\tau=0$ and passes through the center of the spacetime, while the regions $z(D_A)$ and $z(D_{\bar{A}})$ do not reach the center.}

\subsection{The wedge of minimal-area extremal surfaces $w(D_A)$.}

Based on these observations, and the observation of Hubeny that the extremal surfaces probe deepest into the bulk in various examples, it is natural to define a second candidate for the region $R(A)$ based on extremal surfaces.

The surface $W(A)$ calculates the entanglement entropy associated with the entire domain of dependence $D_A$ (equivalently, the largest spacelike surface in $D_A$). We can also consider the entanglement entropy associated with any smaller causal development region within $D_A$. For any such region $C$, there will be an associated surface $W(C)$ (as defined above) whose area computes the entanglement entropy (according to the proposal). Define a bulk region $w(D_A)$ as the set of all points contained on some minimal-area\footnote{Here, we mean minimal area among the set of extremal surfaces with the same boundary.} extremal codimension-two surface whose boundary coincides with the boundary of a spacelike codimension-one region in $D_A$. The area of each such codimension-two surface is (according to \cite{Hubeny:2007xt}) equal to the entanglement entropy of the corresponding boundary region. Thus, the region  $w(D_A)$ directly corresponds to the region of the bulk whose geometry is probed by entanglement observables. As we have seen, the region $w(D_A)$ generally extends beyond the region $z(D_A)$.

From the region $w(D_A)$, we can define a larger region $\hat{w}(D_A)$ as the domain of dependence of the region $w(D_A)$. As discussed above, knowing the geometry (and other fields) in $w(D_A)$ and the bulk gravitational equations should allow us to reconstruct the geometry in $\hat{w}(D_A)$.

We would now like to understand whether the region $\hat{w}(D_A)$ obeys the constraints outlined above. Constraints 1 and 2 are satisfied by definition. It is straightforward to show that Constraint 3 is satisfied assuming that the following conjecture holds:
\vskip 0.1 in
\noindent
{\bf Conjecture C1:}  {\it If $D_A$ and $D_B$ are domains of dependence for non-intersecting regions $A$ and $B$ of a spacelike slice of the boundary spacetime, then $w(D_A)$ and $w(D_B)$ are spacelike separated.}
\vskip 0.1 in
\noindent
Supposing that this holds, if $p$ is in $J(\hat{w}(D_B))$, then there exists a causal curve through $p$ intersecting $\hat{w}(D_B)$, and by definition of $\hat{w}$, this causal curve also intersects $w(D_B)$. If $p$ is also in $\hat{w}(D_A)$, then every causal curve through $p$ intersects $w(D_A)$. Thus, there exists a causal curve that intersects both $w(D_B)$ and $w(D_A)$, which violates ${\bf C1}$. We conclude that  $\hat{w}(D_A)$ satisfies Constraints 1, 2 and 3 assuming that Conjecture ${\bf C1}$ holds.

\subsubsection*{Aside: proving Conjecture C1}

While a proof (or refutation) of Conjecture ${\bf C1}$ is left to future work, we make a few additional comments here.

For the case of static spacetimes, it is straightforward to prove a result similar to ${\bf C1}$.
\vskip 0.1 in
\noindent
{\it Let  $A_1$ and $A_2$ be two non-intersecting regions of the $t=0$
boundary slice of a static spacetime, with $B_1$ and $B_2$ spacelike regions
in $A_1$ and $A_2$, respectively.  Let $W(B_1)$ and $W(B_2)$ be the minimal
surfaces in the $t=0$ slice of the bulk spacetime with
$\partial W(B_1) = \partial B_1$ and $\partial W(B_2) = \partial B_2$.
Then $W(B_1)$ and $W(B_2)$ cannot intersect.}
\vskip 0.1 in
\noindent
To show this, consider the part of $W(B_1)$ contained in the region of the $t=0$ slice bounded by $W(B_2)$ and $B_2$, and the part of $W(B_2)$ contained in the region of the $t=0$ slice bounded by $W(B_1)$ and $B_1$. If these two pieces have different areas, then by swapping the two pieces, either the new surface $W(B_1)$ or the new surface $W(B_2)$ will have a smaller area than before, contradicting the assumption that these were minimal-area surfaces. If the two pieces have the same area, the modified surfaces will have the same area as before, but the new surfaces will be cuspy\footnote{The surfaces $W(B_1)$ and $W(B_2)$ cannot be tangent at their intersection because there should be a unique extremal surface passing through a given point with a specified tangent plane to the surface at this point.}, such that we can decrease the area by smoothing the cusps.

In attempting a more general proof, it may be useful to note that Conjecture {\bf {C1}} is equivalent to the following statement (with some mild assumptions):
\\
\\
{\bf{Conjecture C2}}: {\it For any spacelike boundary region $C$, the surface $W(C)$ is spacelike separated from the rest of $w(D_C)$.}
\\
\\
To see the equivalence, assume first that {\bf {C1}} holds and let $A=C$ and $B=\bar C$.  If we assume
the generic case that $W(C)$ is the same as $W(\bar C)$, then $W(C) =
W(B) \subset w(D_B)$ must be spacelike separated from $w(D_A)=w(D_C)$.
Conversely, for two disjoint regions $A$ and $B$, let $C$ be any region such
that $A \subset C$ and $B \subset \bar C$.  By definition, we have
that  $w(D_A)\subset w(D_C)$ and  $w(D_B)\subset w(D_{\bar C})$.
Assuming again that  $W(C) = W(\bar C)$, Conjecture {\bf {C2}}
implies that there is a spacelike path connecting any point
in $w(D_A)\subset w(D_C)$ with any point $p$ in $W(C)$,
and that there also exists a spacelike path connecting any point
in $w(D_B)\subset w(D_{\bar C})$ with the same point $p$.
Therefore, there is a spacelike path (through $p$) connecting any point in
$w(D_A)$ with any point in $w(D_B)$, as required for {\bf {C1}}.

While {\bf {C1}} is immediately more useful,  {\bf {C2}}  might be
easier to prove.  Consider any boundary region $C$ and any point
$p$ in $w(D_C)$.  Then there exists a spacelike
codimension-one region $I_p$ in the domain of dependence $D_C$
such that $p \in W(I_p)$.  $I_p$ can be extended to
a spacelike surface $A_I$ homologous with $C$, with the same boundary
as $C$, $\delta_{A_I} = \delta_C$.  The surface which calculates
entanglement entropy is the same for $A_I$ and $C$: $W(A_I)=W(C)$.
Consider now a one-parameter family of surfaces $S(\lambda)$, which continuously
interpolate between $A_I=S(0)$ and $I_p=S(1)$, and the corresponding
family of bulk minimal surfaces $W(S(\lambda))$ interpolating between
$W(C)$ and $W(I_p)$.  It is plausible that these bulk minimal surfaces
change smoothly and that their deformations are spacelike;
following the flow, we can find a
spacelike path from $p$ to $W(C)$, which would complete the proof
of the Conjecture  {\bf {C2}}.

We leave further investigation of the general validity of ${\bf C1}$ as a question for future work.\footnote{We note here that the restriction to minimal extremal surfaces (rather than all extremal surfaces) is essential for the validity of this conjecture. In static spacetimes with metric of the form $ds^2 = -f(r)dt^2 + dr^2/g(r) + r^2 d \Omega^2$ where $g(0)=1$ and $g(r) \to r^2$, it is possible that extremal surfaces bounded on one hemisphere intersect extremal surfaces bounded on the other hemisphere in cases where $g(r)$ is not monotonically increasing. For these examples, ${\bf C1}$ would fail if the definition of $w$ did not restrict to minimal surfaces.}

\subsubsection*{Possible connection between the geometry of $W(A)$ and the spectrum of $\rho_A$}

To summarize the discussion so far, the region $\hat{w}(D_A)$ satisfies conditions 1, 2 and 3 assuming that Conjecture ${\bf C1}$ is correct. Thus, $\hat w(D_A)$ is a possible candidate for the region $R(A)$. A rather nice feature of this possibility is that $\hat{w}(D_A)$ intersects $\hat{w}(D_{\bar{A}})$ along the codimension-two surface $W(A)=W(\bar{A})$ defined above. Thus, the surface $W$ represents the information in the bulk common to $\hat{w}(D_A)$ and $\hat{w}(D_{\bar{A}})$. The area of this surface corresponds to the von Neumann entropy of $\rho_A$, which is the simplest information shared by $\rho_A$ and $\rho_{\bar{A}}$. We might then conjecture that the full spectrum of $\rho_A$ (which is the same as the spectrum of $\rho_{\bar{A}}$ and represents the largest set of information common to $\rho_A$ and $\rho_{\bar{A}}$) encodes the full geometry of the surface $W$ (i.e. the largest set of information common to $\hat{w}(D_A)$ and $\hat{w}(D_{\bar{A}})$).

\subsubsection*{Reconstructing bulk metrics from extremal surface areas}

Before proceeding, let us ask whether it is even possible that the areas of extremal surfaces with boundary in some region $D_A$ carry enough information to reconstruct the geometry in $w(D_A)$.

Consider the simple case of a 1+1 dimensional CFT on a cylinder with $D_A$ a diamond-shaped region on the boundary. Given any state for the CFT, we could in principle compute the entanglement entropy associated with any smaller diamond-shaped region bounded by the past lightcone of some point in $D_A$ and the forward lightcone of some other point. This would give us one function of four variables, since each of the two points defining the smaller diamond-shaped region is labeled by two coordinates. Assuming the state has a geometrical bulk dual description, the bulk geometry will be described by a metric which consists of several functions of three variables.\footnote{We are ignoring the possible extra compact dimensions in the bulk.} These functions allow us to determine the entanglement entropy from the geometry in the wedge $w(D_A)$ via the Takayanagi et. al. proposal, so we have a map from the space of metrics to the space of entropy functions. Small changes in the geometry of the wedge $w(D_A)$ will generally affect the areas of some of the minimal surfaces, while small changes in the geometry outside the wedge will generally not affect these areas. It is at least plausible that the entanglement information could be used to fully reconstruct the geometry in the wedge in some cases, since the map from wedge geometries into the entanglement information is a map from finitely many functions of three variables to a function of four variables, and it is possible for such a map to be an injection.

A proven result of this form in the mathematics literature \cite{pu} is that for two-dimensional {\it simple}\footnote{See \cite{pu} for the definition of a simple manifold.} compact Riemannian manifolds with boundary, the bulk geometry is completely fixed by the distance function $d(x,y)$ between points on the boundary (the lengths of the shortest geodesics connecting various points). This implies that for static three-dimensional spacetimes, the spatial metric of the bulk constant time slices can be reconstructed in principle if the entanglement entropy is known for arbitrary subsets of the boundary. However, we are not aware of any results about the portion of a space that can be reconstructed if the distance function is known only on a subset of the boundary, or of any results that apply to Lorentzian spacetimes.

\subsubsection*{Cases when $R(A)$ cannot be larger than $\hat{w}(D_A)$}

We saw above that in special cases, $z(D_A)$ together with $J(z(D_{\bar{A}}))$ cover the entire spacetime, so Constraint 3 is just barely satisfied for $z$ (or $\hat{z}$). For these examples, if $z(D_{\bar{A}})$ is in $R(\bar{A})$ then $R(A)$ cannot possibly be larger than $z(D_A)$. On the other hand, for generic spacetimes, we argued that only a portion of the spacetime is covered by $z(D_A)$ and  $J(z(D_{\bar{A}}))$, leaving the possibility that $R(A)$ could be larger than $z(D_A)$. In these examples, extremal surfaces from $A$ typically extend into the region not covered by $z(D_A)$ or $z(D_{\bar{A}})$ (or the causal past/future of these), and this motivated us to consider $\hat{w}(D_A)$ as a larger possibility for $R(A)$.

We will now see that in a much wider class of examples, $\hat{w}(D_A)$ together with $J(w(D_{\bar{A}}))$ do cover the entire spacetime. To see this, recall that the surfaces $W(A)$ and $W(\bar{A})$ computing the entanglement entropy of the entire regions $A$ and $\bar{A}$ are the same by definition, as long as $A$ and $\bar{A}$ are homologous in the bulk.\footnote{The only possible exception would be the case where there are two extremal surfaces with equal area having boundary $\partial A$. In this case, we might call one $W(A)$ and the other $W(\bar{A})$.} Now, suppose that for a one-parameter family of boundary regions $B(\lambda) \subset A$ interpolating between $A$ and a point (assuming $A$ is contractible),  the surfaces $W(B(\lambda))$ change smoothly. Similarly, suppose that for a one-parameter family of boundary regions $B'(\lambda) \subset \bar{A}$ interpolating between $\bar{A}$ and a point (assuming $\bar{A}$ is contractible),  the surfaces $W(B'(\lambda))$ change smoothly. Then the union of all surfaces $W(B(\lambda))$ and $W(B'(\lambda))$ covers an entire slice of the bulk spacetime. In this case, for any point $p$ in the bulk spacetime, either there is a causal curve through $p$ that intersects $\cup_\lambda W(B(\lambda)) \subset w(D_A)$ or else every causal curve through $p$ intersects $\cup_{\lambda} W(B'(\lambda)) \subset w(D_{\bar{A}})$. This shows that $\hat{w}(D_A)$ together with $J(w(D_{\bar{A}}))$ cover the entire spacetime.

To summarize, in cases where $W(B)$ varies smoothly with $B$ as described above, we have that $\hat{w}(D_A)$ together with $J(w(D_{\bar{A}}))$ cover the entire spacetime. Thus, by Constraint 3, {\it with this smoothness condition}, if $\hat{w}(\bar{A}) \subset R(\bar{A})$ then $R(A)$ cannot be larger than $\hat{w}(D_A)$.\footnote{An alternative condition that leads to the same conclusion is that $w(D_A) \cup w(D_{\bar{A}})$ includes a Cauchy surface.} While there are many examples of spacetimes for which this smooth variation does not occur (e.g. as described in the next section), spacetimes satisfying the condition are not particularly special.

\subsubsection*{An example where $R(A)$ is strictly larger than $\hat{w}(D_A)$ }

We have seen that $\hat{w}(D_A)$ is in some sense a maximally optimistic proposal for $R(A)$ in cases where a particular smoothness condition is satisfied or when $w(D_A) \cup w(D_{\bar{A}})$ includes a Cauchy surface. We will now see that these conditions can fail to be true in some cases, and that in these cases,  $R(A)$ must be larger than $\hat{w}(D_A)$ for some choice of $A$.

\begin{figure}
\centering
\includegraphics[width=0.7\textwidth]{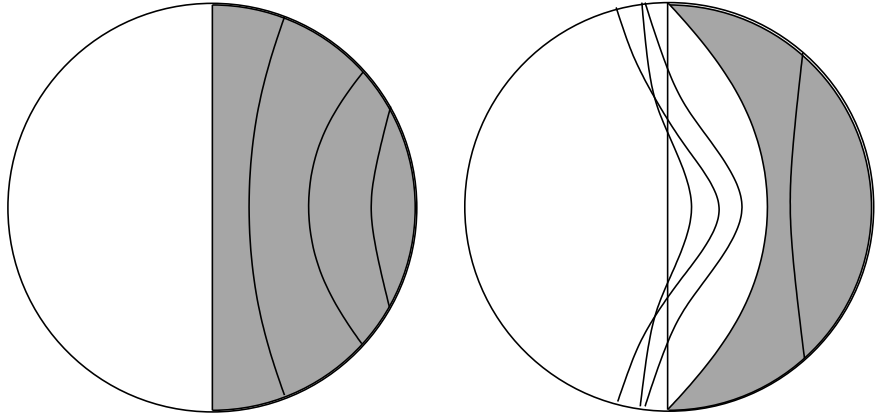}
\caption{Different possible behaviors of extremal surfaces in spherically symmetric static spacetimes. Shaded region indicates $w(D_A)$ where $A$ is the right hemisphere. The boundary of the shaded region on the interior of the spacetime is the minimal area extremal surface bounded by the equatorial $S^{d-1}$.}
\label{extremals}
\end{figure}

Consider the simple example of static spherically symmetric spacetimes with metric of the form $ds^2 = -f(r)dt^2 + dr^2/g(r) + r^2 d \theta^2$ where $g(0)=1$ and $g(r) \to r^2$ for large $r$. For any spacetime of this form, the extremal codimension-two surfaces bounded by spherical regions on the boundary will be constant-time surfaces in the bulk that can easily be computed. By symmetry, there always exists an extremal surface through the center of the spacetime whose boundary is an equatorial $S^{d-1}$ of the boundary $S^d$. Now, moving out towards the boundary along some radial geodesic, there will be a unique extremal surface passing through each point and normal to the radial line.

In some cases (e.g. pure AdS), the boundary spheres for these extremal surfaces shrink monotonically as we approach the boundary, as shown in the left half of Figure~\ref{extremals}. However, there are other cases for which $g(r)$ is not monotonic where the extremal surfaces shrink in the opposite direction, then grow, then shrink again, as shown in the right half of Figure~\ref{extremals}.\footnote{As an explicit example, we have considered the case of a charged massive scalar field coupled to gravity, with scalar field of the form $\phi(r) = e^{i \omega t} f(r)$. Spherically-symmetric configurations of this type with non-zero charge are known as ``boson-stars'' \cite{Astefanesei:2003qy}. We find that for fixed $\psi(0)$, the metric function $g(r)$ is monotonically increasing for sufficiently small values of the scalar field mass, while for sufficiently large values we can have the behavior shown on the right in Figure~\ref{extremals}.} In these cases, boundary spheres with angular radius in a neighborhood of $\pi/2$ will bound multiple extremal surfaces in the bulk. The extremal surface of minimum area in these cases is always one that is contained within one half of the bulk space (otherwise we could construct intersecting minimal surfaces bounding disjoint regions of the boundary). Considering only the minimal surfaces, we find that there exists a spherical region in the middle of the spacetime penetrated by no such surface. Thus, even if we choose $D_A$ to be the entire spacetime boundary, the region $w(D_A)$ excludes the region $r<r_0$ for some $r_0$. In this case, we have all information about the field theory (assumed to be a pure state), so $R(A)$ should be the entire spacetime.

\begin{figure}
\centering
\includegraphics[width=0.4\textwidth]{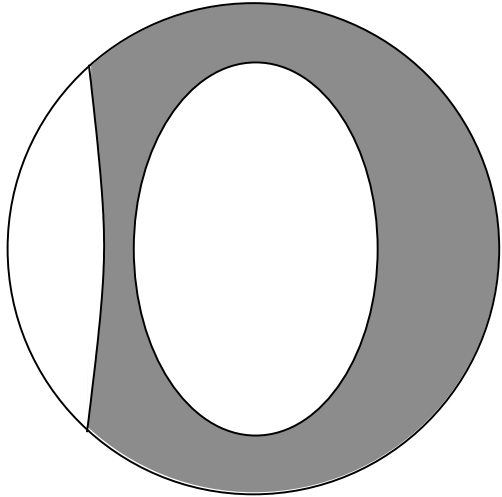}
\caption{Region $w(D_A)$ (shaded) where $A$ is a boundary sphere of angular size greater than $\pi$. No minimal surface with boundary in $A$ penetrates the unshaded middle region.}
\label{hole}
\end{figure}

More generally, the region $w(D_A)$ in these cases will have a ``hole'' if $A$ is chosen to be any boundary sphere with angular radius between $\pi/2$ and $\pi$, as shown in Figure~\ref{hole}. Note, however, that the central region is included in $z(D_A)$ for sufficiently large $A$, so $z(D_A) \not\subset w(D_A)$ in these cases.

\section{Discussion}

In this note, we have presented various consistency constraints on the region $R(A)$ of spacetime which can in principle be reconstructed from the density matrix $\rho_A$ for a spatial region $A$ of the boundary with domain of dependence $D_A$. We have argued that the $z(D_A) \equiv J^+(D_A) \cap J^-(D_A)$ and its domain of dependence $\hat{z}(D_A)$ should be contained in $R(A)$ and that $\hat{z}(D_A)$ satisfies our consistency constraints. Since entanglement observables calculated from $\rho_A$ correspond to extremal surfaces that typically probe a region of spacetime beyond $\hat{z}(D_A)$, we have also considered the union of these surfaces $w(D_A)$ and its domain of dependence $\hat{w}(D_A)$ as a possibility for $R(A)$ that is often larger than $\hat{z}(D_A)$. We have seen that $\hat{w}(D_A)$ also satisfies our constraints (assuming Conjecture ${\bf C1}$), and that if $\hat{w}(D_A) \subset R(A)$ generally, then $R(A) = \hat{w}(D_A)$ for a broad class of spacetimes.

\subsubsection*{A false constraint}

The constraints discussed in this note are essentially consistency requirements that do not make use of details of the AdS/CFT correspondence. It is interesting to ask whether there exist any more detailed conditions that could constrain the region $R(A)$ further.

\begin{figure}
\centering
\includegraphics[width=0.15\textwidth]{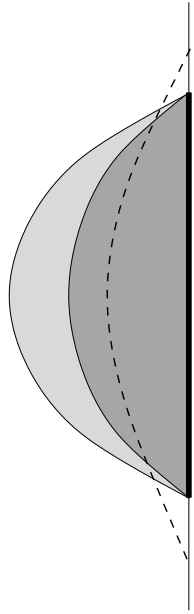}
\caption{Spatial $t=0$ slice of $w(D_A)$ (light shaded plus dark shaded) and $z(D_A)$ (dark shaded) for a planar AdS black hole. The dashed curve is a spatial geodesic with endpoints in $\bar{A}$. Knowledge of observables obtained from $\rho_{\bar{A}}$ alone allow us to compute the length of this geodesic.}
\label{g-overlap}
\end{figure}

It may be instructive to point out a somewhat plausible constraint that turns out to be false. For two non-intersecting regions $A$ and $B$ of the boundary spacetime, it may seem that the region $G(B)$ of the spacetime used to construct field theory observables in $B$ should not intersect the region $R(A)$ dual to the density matrix $\rho_A$. The argument might be that if the physics in $R(A)$ is the bulk manifestation of information in $\rho_A$, we cannot expect to learn anything about this region knowing only $\rho_B$. It would seem that this would be telling us directly about $\rho_A$ knowing only $\rho_B$. Perhaps surprisingly, it is easy to find an example where neither $w(D_A)$ nor $z(D_A)$ satisfies this constraint, see Figure~\ref{g-overlap}.

In the planar AdS black hole geometry, take the region $A$ to be a ball-shaped region on the boundary. In this case, it is straightforward to check that spatial geodesics with endpoints in $\bar{A}$ intersect both $w(D_A)$ and $z(D_A)$. Thus, the constraint $R(A) \cap G(\bar{A}) = \emptyset$ can't be correct if $z(D_A) \subset R(A)$. In hindsight, it is not difficult to understand the reason. Knowledge of the density matrix $\rho_{\bar{A}}$ allows us to reconstruct $R(\bar{A})$. There could be many states of the full theory that give rise to the same density matrix $\rho_{\bar{A}}$. For any such state with a classical gravity dual description, the dual spacetime geometry must be such that spatial geodesics anchored in $D_{\bar{A}}$ have the same lengths as in the original spacetime we were considering. But there can be many such spacetimes. So using the information in $\rho_{\bar{A}}$, we are not learning directly about $\rho_A$, only about the family of density matrices $\rho_A^\alpha$ such that the pair $(\rho_{\bar{A}}, \rho_A^\alpha)$ can arise from a pure state $|\Psi \rangle$ that has a geometrical gravity dual.\footnote{Don Marolf has pointed out to us that the connection between two-point functions and the lengths of spatial geodesics has been argued to fail for spacetimes that do not satisfy certain analyticity properties \cite{marolf}. It is likely that demanding such properties imposes even stronger constraints connecting $\rho_A$ and $\rho_{\bar{A}}$.}

\subsubsection*{Spacetime emergence and entanglement}

\begin{figure}
\centering
\includegraphics[width=0.9\textwidth]{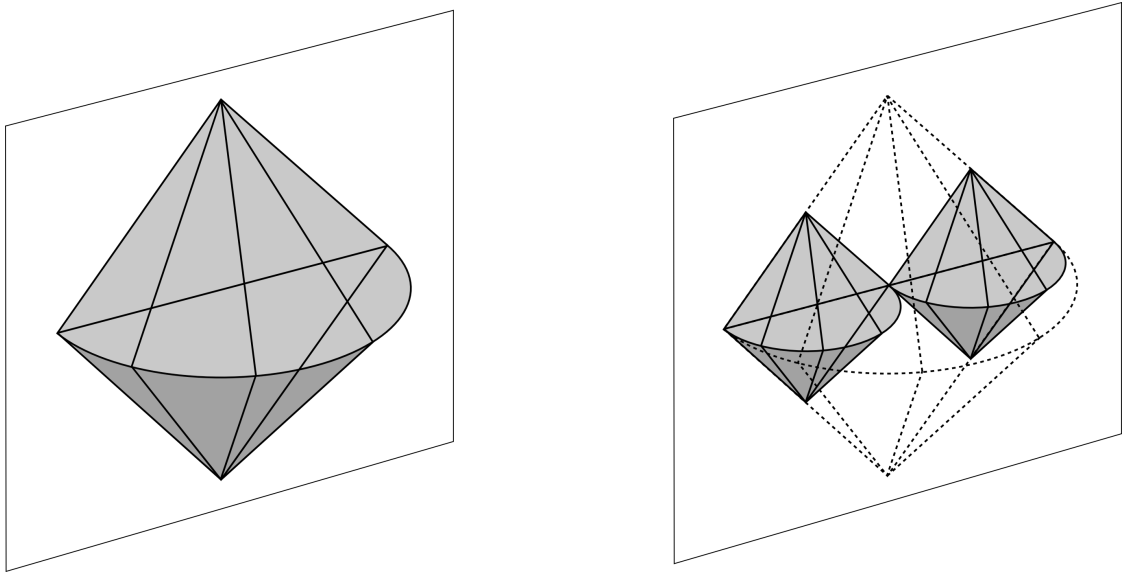}
\caption{The region of spacetime reconstructible from density matrices $\rho_B$ and $\rho_C$ (shaded, right hand side picture) is smaller than that reconstructible from $\rho_{B \cup C}$ (shaded, left hand side picture). Reconstruction of $R(B \cup C) - (R(B)\cup R(C))$ (interior of dotted frame outside of the two shaded regions) requires knowledge of entanglement between degrees of freedom in $B$ and $C$.}
\label{j1}
\end{figure}

\begin{figure}
\centering
\includegraphics[width=0.9\textwidth]{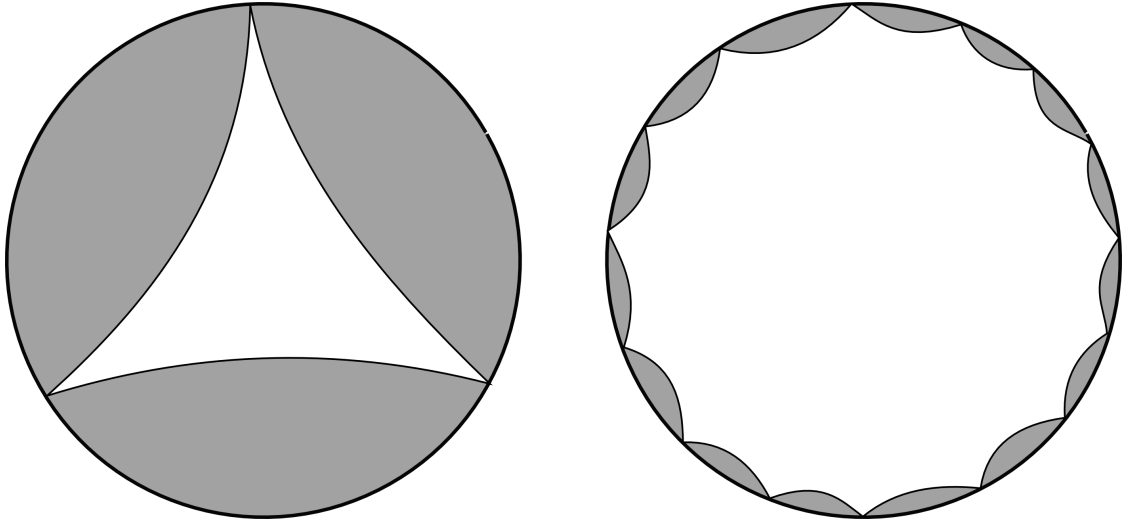}
\caption{The region of spacetime reconstructible from density matrices $\rho_{A_i}$ lies arbitrarily close to the boundary (illustrated here on a spatial slice). The ability to reconstruct the bulk geometry depends entirely on the knowledge of entanglement among the various boundary regions.}
\label{wianek}
\end{figure}

The observations in this note highlight the importance of entanglement in the emergence of the dual spacetime. Consider a collection $\{A_i\}$ of subsets on the boundary such that $\cup A_i$ covers an entire boundary Cauchy surface. In a classical system, knowing the configuration and time derivatives of the fields in each of these regions would give us complete information about the physical system. Quantum mechanically, however, complete information about the system consists of two ingredients: (i) the density matrices $\rho_{A_i}$, and (ii) the entanglement between the various regions.

If we subdivide a set $A \to \{B,C\}$ and pass from $\rho_A \to \{\rho_B,\rho_C\}$, we lose information about the entanglement between $B$ and $C$. In the bulk picture, the region of spacetime that we can reconstruct (for any $R$ satisfying our constraints) is significantly smaller than before, as we see in Figure~\ref{j1}. The region of spacetime that we can no longer reconstruct corresponds to the information about the entanglement between the degrees of freedom in $B$ and $C$ that we lost when subdividing.

As we divide the boundary into smaller and smaller sets $A_i$, we retain information about entanglement only at successively smaller scales, while the bulk space $\cup R(A_i)$ that can be reconstructed retreats ever closer to the boundary (Figure \ref{wianek}). Conversely, knowledge of the bulk geometry at successively greater distance from the boundary requires knowledge of entanglement at successively longer scales.\footnote{A very similar picture was advocated in \cite{swingle}.} In the limit where $A_i$ become arbitrarily small, we know nothing about the bulk spacetime even if we know the precise state for each of the individual degrees of freedom via the matrices $\rho_{A_i}$.
In this sense, the bulk spacetime is entirely encoded in the entanglement of the boundary degrees of freedom.

\section*{Acknowledgments}

We are especially grateful to Veronika Hubeny and Mukund Rangamani for important comments and helpful discussions. This work is supported in part by the Natural Sciences and Engineering Research Council of Canada and by the Canada Research Chairs Programme.

\end{document}